\documentclass[aps,pre,amsmath,amssymb,reprint]{revtex4-1}
\usepackage{graphicx}% Include figure files
\usepackage{dcolumn}% Align table columns on decimal point
\usepackage{bm}% bold math
\usepackage{amsfonts}
\usepackage{latexsym}
        \def\refeq#1{Eq.~(\ref{eq:#1})}
        \def\reffig#1{Fig.~\ref{fig:#1}}
        \def\refsec#1{Section~\ref{sec:#1}}

        \def\A0{A_0}

	\def\adj{\dagger}
	\def\dirmod{\psi}
	\def\adjmod{\psi^\adj}
	\def\linop{\mathcal{L}}
	\def\eigval{\lambda}
	\def\pertamp{\varepsilon}

	\def\glvar{\phi}

	\def\bra{\langle}
	\def\ket{\rangle}

	\def\gammaeff{\gamma_{eff}}

\begin{document}

\preprint{???}

% \title[Optimal perturbations of non-parallel wakes and their stabilizing effect]
%       {Optimal perturbations of non-parallel wakes and their stabilizing effect on the global instability}

\title[]{On the stabilizing mechanism of 2D absolute and global instabilities by 3D streaks}

\author{Carlo Cossu}
\affiliation{Institut de M\'ecanique des Fluides de Toulouse (IMFT), CNRS and Universit\'e de Toulouse\\ 
All\'ee du Professeur Camille Soula, 31400 Toulouse, France}
\altaffiliation[Also at ]{ D\'epartement de M\'ecanique, \'Ecole Polytechnique, 91128 Palaiseau, France}

% \date{\today}% It is always \today, today,
\date{April 4, 2014}% It is always \today, today,
             %  but any date may be explicitly specified

\begin{abstract}
Global and local absolute instabilities of 2D wakes are known to be stabilized by spanwise periodic modulations of the wake profile.
The present study shows that this stabilizing effect is of general nature and can be mimicked by enforcing spanwise periodic modulations of the wave advection velocity in the generalized complex Ginzburg-Landau equation.
The first order sensitivity of the absolute and global growth rate to the enforced modulation is zero, exactly as in the Navier-Stokes case. 
We show that a second order sensitivity analysis is effective to quantify and interpret the observed stabilizing effect.
The global growth rates predicted by the second order expansion closely match those issued from a direct computation of the eigenvalues.
It is shown that, at leading order, the modulation of the wave advection velocity alters the effective wave diffusion coefficient in the dispersion relation and that this variation induces a reduction of the  absolute growth rate and the stabilization of the global instability.
\end{abstract}

%Valid PACS numbers may be entered using the \verb+\pacs{#1}+ command.
                             % Classification Scheme.
%Use showkeys class option if keyword
                              %display desired
\maketitle

%%%%%%%%%%%%%%%%%%%%%%%%%%%%%%%%%%%%%%%%%%%%%%%%%%%%%%%%%%%%%%%%%%%%%%%%%%%%%%%
\section{Introduction}

% CITE AND DISCUSS RALLISON PAPER, MENTION HINCH SUGGESTION?

Robust self-sustained vortex shedding is present in bluff body wakes above the critical Reynolds number where a global linear instability sets in.
The global instability is supported by a finite region of local absolute instability in the near wake \cite{Chomaz1988,Monkewitz1988,Huerre1990}. 
Vortex shedding being associated to undesired effects, such as noise, increased drag, buffetting and unsteady loads, its control is the object of continued interest.

It has long been known that vortex shedding can be attenuated and even suppressed by forcing spanwise periodic modulation of the wake \cite{Choi2008}. 
These 3D modulations of the 2D unstable wake have been typically obtained by wrapping an helical cable around a cylinder \citep[e.g.][]{Zdravkovich1981} or designing spanwise periodic trailing and/or leading  edges of the bluff body \cite{Tanner1972,Tombazis1997,Bearman1998,Darekar2001} or with spanwise periodic blowing and suction \cite{Kim2005}.
The stabilization mechanism is, to this date, understood as an inhibition of vortex roll-up redistribution induced by the 3D redistribution of the shear layers \cite{Darekar2001,Choi2008} with vortex tilting cited as a key process \cite{Hwang2013}.
Recently, progress has been made in relating the stabilizing effect of 3D control to the basic instability mechanisms underlying the global instability.
It has first been shown that appropriate 3D spanwise periodic perturbations of absolutely unstable 2D parallel wakes, lead to a reduction of the absolute growth rate \cite{Hwang2013} and, for sufficiently large forcing amplitudes, to the suppression of the absolute instability \cite{DelGuercio2014}. 
It has then been shown that optimal 3D perturbations of non-parallel wakes decrease the global mode growth rate up to its complete stabilization for sufficiently large forcing amplitudes \cite{DelGuercio2014b,DelGuercio2014c}.

Despite this recent progress, the core mechanism underlying the stabilizing effect of 3D perturbations on the absolute instability has not been completely elucidated from an hydrodynamic stability perspective. 
%beyond the interpretation in terms of vortex dynamics.
For instance, it is not clear if the stabilizing mechanism is limited to the particular vortex dynamics in wake flows or is limited to the physics of Navier-Stokes equations or if it is of more general nature. 
In this study we investigate the nature of the stabilizing mechanism by mimicking it in the generalized complex Ginzburg-Landau equation. This equation, which describes the wave amplitude evolution in a bifurcating open flow \cite{Newell1969,Stewartson1971}, arises in many different areas of theoretical physics \cite{Aranson2002} and has already been used to illustrate the relation between the local absolute and global instabilities \cite{Chomaz1988,Huerre1990} or the relation between local convective instabilities and global streamwise non-normality \cite{Cossu1997c,Chomaz2005} in open shear flows.
This equation has also been used to discuss the 2D control of wake flows \cite{Roussopoulos1996,Lauga2003,Lauga2004,Aamo2005}.

One peculiar characteristic of spanwise periodic modification of 2D wake flows is that the usual first order structural sensitivity  \cite{Bottaro2003,Chomaz2005} of the absolute growth rate and of the global mode growth rate to spanwise periodic control is found to be zero \cite{Hwang2006,Hwang2013,DelGuercio2014,DelGuercio2014b,DelGuercio2014c}. 
The same is found when considering the generalized complex Ginzburg-Landau equation.
A second  order sensitivity analysis is therefore introduced in \refsec{form} in order to assess {\it a priori} and theoretically analyse the stabilizing effect of spanwise periodic structural perturbations of the Ginzburg-Landau operator.
We anticipate that spanwise periodic modulations of the wave advection velocity can reproduce the stabilizing effect of spanwise periodic perturbations of 2D wakes wakes.
In particular, the influence of spanwise modulations of the advection velocity is assessed for the  global instability of a generalized non-parallel complex Ginzburg-Landau equation and for the local absolute instability in the parallel case, as detailed in  \refsec{Results}. 
The main results of the analysis are then summarized and discussed in \refsec{Concl}.
Details of the numerical computations are given in Appendix~\ref{sec:Num}.

% The paper is organized as follows.
% The second order structural sensitivity analysis is summarized in \refsec{form}.

% Using the second order sensitivity results, we will also show that in the parallel case, the leading effect of the spanwise modulations of the advection velocity is to alter the wave diffusivity and that this effect results in the stabilization of the absolute growth rate if the imaginary part of the diffusivity is larger than the real part.
%  
% COMPLETE HERE

%%%%%%%%%%%%%%%%%%%%%%%%%%%%%%%%%%%%%%%%%%%%%%%%%%%%%%%%%%%%%%%%%%%%%%%%%%%%%%%
\section{Problem formulation}
\label{sec:form}

%------------------------------------------------------------------------------
\subsection{Second order structural sensitivity of eigenvalues of a linear operator}
\label{sec:Sens2}

The usual first order structural sensitivity analysis of the eigenvalues to perturbations of the linear operator \cite{Bottaro2003,Chomaz2005} can be extended to second order in order to analyze problems where the first order sensitivity is zero.
Let us consider the eigenvalue problem:
\begin{equation} \eigval \dirmod = \linop \dirmod, 
\label{eq:eigprob}
\end{equation}
where the linear operator $\linop$ is given by a reference linear operator $\linop_0$ perturbed with $\pertamp \linop_1$:
\begin{equation} \linop = \linop_0 + \pertamp \linop_1. \end{equation}
In the limit of infinitesimal perturbation amplitudes $\pertamp$, the eigenvalue  $\eigval$ and the corresponding eigenfunction  $\dirmod$ can be sought as:
\begin{eqnarray} 
\eigval &=& \eigval_0 + \pertamp \eigval_1 + \pertamp^2 \eigval_2 \\
\dirmod &=& \dirmod_0 + \pertamp \dirmod_1 + \pertamp^2 \dirmod_2.
\end{eqnarray}
which replaced in \refeq{eigprob} provides the set of equations to be satisfied at zero-th, first and second order respectively:
\begin{eqnarray} 
\eigval_0 \dirmod_0 &=& \linop_0 \dirmod_0 \\
\label{eq:ord1}
\eigval_0 \dirmod_1 + \eigval_1 \dirmod_0 &=& \linop_0 \dirmod_1 + \linop_1 \dirmod_0 \\
\label{eq:ord2}
\eigval_0 \dirmod_2 + \eigval_1 \dirmod_1 + \eigval_2 \dirmod_0 &=& \linop_0 \dirmod_2 + \linop_1 \dirmod_1. 
\end{eqnarray}
Denoting by $\linop^\adj_0$ the adjoint of $\linop_0$ with respect to the inner product $\bra , \ket$, to each eigenvalue $\eigval_0$ of $\linop_0$ corresponds an eigenvalue of $\linop^\adj_0$ which simply is the complex conjugate of $\eigval_0$, whose corresponding eigenfunction is denoted by $\adjmod_0$.
The first order sensitivity $\eigval_1$ is easily retrieved \cite{Bottaro2003,Chomaz2005} by projecting \refeq{ord1} on $\adjmod_0$
and making use of the identity $\bra \adjmod_0 , \eigval_0 \dirmod_1 \ket =  \bra \adjmod_0 , \linop_0 \dirmod_1 \ket$:
\begin{equation} 
\eigval_1 = \frac{ \bra \adjmod_0 , \linop_1 \dirmod_0 \ket }{  \bra \adjmod_0 , \dirmod_0 \ket }. 
\label{eq:sens1}
\end{equation}
We are especially interested in the case where the first order sensitivity is zero ($\eigval_1 = 0$), a situation which is e.g. encountered when spanwise uniform basic flows are perturbed with spanwise periodic perturbations \cite{Hwang2006,Hwang2013,DelGuercio2014,DelGuercio2014b}.
In this case, an expression for the second-order sensitivity $\lambda_2$ can be retrieved by projecting the second order \refeq{ord2} on $\adjmod_0$ and making use of the identity $\bra \adjmod_0 , \eigval_0 \dirmod_2 \ket =  \bra \adjmod_0 , \linop_0 \dirmod_2 \ket$:
\begin{equation} 
\eigval_2 = \frac{ \bra \adjmod_0 , \linop_1 \dirmod_1 \ket }{  \bra \adjmod_0 , \dirmod_0 \ket } ,
\label{eq:sens2}
\end{equation}
The first order eigenfunction correction $\dirmod_1$ is retrieved by solving \refeq{ord1} which, as $\eigval_1=0$, reduces to:
\begin{equation}  
\left( \linop_0 - \eigval_0 \right) \dirmod_1 = - \linop_1 \dirmod_0. 
\label{eq:psi1eq}
\end{equation}
This linear system is singular, as the adjoint homogeneous problem admits the non trivial solution $\adjmod_0$. 
However, as it is assumed that $\eigval_1=0$, then from \refeq{sens1} follows that  $\bra \adjmod_0 , \linop_1 \dirmod_0 \ket = 0$ and therefore the right hand side of \refeq{psi1eq} is orthogonal to the non-trivial kernel of the adjoint problem. 
In such a case Fredholm theorems \cite{Kreyszig1978} guarantee that solutions to \refeq{psi1eq} exist and are of the type
% \begin{equation}  
$\dirmod_1 = \dirmod_1^\perp + C \dirmod_0$, 
% \label{eq:psi1sol}
% \end{equation}
where $\dirmod_1^\perp$ is the particular solution orthogonal to $\dirmod_0$ and $C$ is an arbitrary constant. 
When $\dirmod_1$, 
%given by \refeq{psi1sol}, 
is replaced in \refeq{sens2}, the term $C \dirmod_0$ is filtered out since its contribution 
${ \bra \adjmod_0 , C \linop_1 \dirmod_0 \ket /  \bra \adjmod_0 , \dirmod_0 \ket } 
= C \eigval_1$ is proportional to $\eigval_1$ which is assumed to be zero.

%------------------------------------------------------------------------------
\subsection{Second order structural sensitivity of the complex Ginzburg-Landau equation}
\label{sec:Sens2CGL}

The generalized linearized complex  Ginzburg-Landau equation $ \partial_t \glvar = \linop \glvar $ describes the evolution of the wave (small) amplitude $\glvar$ of a bifurcating basic state ($\Phi=0$ in this framework) with the linearized operator defined by:
\begin{equation}
%       \partial_t \glvar  =
       {\cal L} = - U \partial_x  +\mu {\mathcal I} + \gamma \partial_{xx} + \partial_{zz},
\label{eq:GLLOP}
\end{equation}
where $U$ is the wave advection velocity, $\mu$ is the local bifurcation parameter and $\gamma=1+ i c_d$ is the streamwise diffusion coefficient. $U$, $\mu$, $c_d$ are assumed real and $\glvar$ is assumed to remain finite $\forall\,x$.
This original complex Ginzburg-Landau equation is here generalized in two different aspects.
The first is to introduce a spatially dependent coefficient $\mu(x)$ to mimic the dynamics of non-parallel flows such as wakes, jets and boundary layers as already done in previous studies \cite{Chomaz1988,Hunt1991,Roussopoulos1996,Cossu1997c,Lauga2003,Aamo2005}. 

The second generalization is related to our goal which is to mimic the passive control of a globally unstable wake obtained by spanwise periodic perturbations of the basic flow \cite{Tanner1972,Bearman1998,Darekar2001,Kim2005,Choi2008,DelGuercio2014,DelGuercio2014b}.
One essential feature of these basic flow modifications is the development of spanwise periodic regions of higher and lower velocity in the unstable region, the so-called streamwise streaks. 
We  assume that the leading effect of these streaks can be reproduced by considering spanwise dependent $U$ and $\glvar$. For physical consistency a spanwise diffusion term is also introduced. 
%This generalized equation must be considered only as a toy model and not as a genuine amplitude equation approximating a real system.
In particular, we set  $ U = U_0 + \pertamp U_1 $, with $ U_1 = U_0 \sin \beta z $ and therefore:
\begin{equation}
 U(z)  = U_0 \left( 1 + \pertamp \sin \beta z \right) 
\label{eq:U3D}
\end{equation}
The amplitude $\pertamp$ represents the streak amplitude and $\beta \neq 0$ is the spanwise wavenumber of the basic flow modulation. At this stage, other probable consequences of the presence of streaks, such as the spanwise modulation of $\mu$ or the streamwise variation of the streak amplitude, are neglected.
With the assumed perturbation of the operator, only through $U_1$, the relevant zero-th and first order operators are
\begin{eqnarray}
\linop_0 &=& - U_0 \partial_x  +\mu(x) {\mathcal I} + \gamma \partial_{xx} + \partial_{zz}, \\
\linop_0^\adj &=&  U_0 \partial_x  + \mu(x) {\mathcal I} + \gamma^* \partial_{xx} + \partial_{zz}, \\
\linop_1 &=& - U_1 \partial_x \equiv - U_0 \sin( \beta z)\,\, \partial_x, 
\label{eq:Ops01}
\end{eqnarray}
where $^*$ denotes the complex conjugate.
% The eigenvalue $\eigval_0$ and eigenfunction $\dirmod_0$ of interest are the most unstable ones, corresponding to $n=1$ in \refeq{2Dspec}.
The reference operator, corresponding to $\pertamp=0$, is therefore `2D' in the sense that all the coefficients and the solutions are assumed independent of $z$, while the perturbed operator is 3D.
The inner product is defined in the standard way as 
$\bra p , q \ket = \int_{x_a}^{x_b} \int_0^{2 \pi / \beta} p^*\, q\, dx\, dz$,
where $x_a$ and $x_b$ are the upstream and downstream boundaries of the control domain.
Considering that $\linop_1 \dirmod_0 = -U_0 \sin ( \beta z) \partial_x \dirmod_0(x)$ and that  $\dirmod_0(x)$ and $\adjmod_0(x)$ do not depend on $z$, it is immediately verified that     
$\bra \adjmod_0 , \linop_1 \dirmod_0 \ket =0 $ and therefore, according to \refeq{sens1}, the first order sensitivity  $\eigval_1$ to spanwise periodic $U$ perturbations is zero. 
This, of course, would  not be the case for spanwise uniform variations of $U$.
We therefore proceed to the second order sensitivity analysis.
The first step of the solution procedure consists in the computation of  $\dirmod_1$ by solving \refeq{psi1eq}, that in the present case is
\begin{equation}  
\left[  \mu - \eigval_0 - U_0 \partial_x  + \gamma \partial_{xx} + \partial_{zz} \right]\, \dirmod_1 = 
U_0 \sin( \beta z) \partial_x \dirmod_0.
\label{eq:GLpsi1eq}
\end{equation}
This inhomogeneous linear system is singular but
the singular and non-singular parts of the equation can be separated by expanding the solution in Fourier sine series in the spanwise direction and then setting to zero the coefficients of all the spanwise harmonics in \refeq{GLpsi1eq}.
It is readily seen that the only non-zero terms of the expansion of $\dirmod_1$ are the spanwise mean and the first harmonic:
$\dirmod_1(x,z) = \overline{\dirmod}_1(x) +  \widetilde{\dirmod}_1(x) \sin( \beta z)$. 
The spanwise uniform and the first harmonic component of \refeq{GLpsi1eq} are:
\begin{eqnarray}  
\label{eq:GLpsi1eqHarm0}
\left[  \mu - \eigval_0 - U_0 \partial_x  + \gamma \partial_{xx}  \right]\, \overline{\dirmod}_1 &=& 0  \\
\left[  \mu - \eigval_0 - \beta^2 - U_0 \partial_x  + \gamma \partial_{xx} \right]\, \widetilde{\dirmod}_1 &=& 
U_0 \partial_x \dirmod_0 
\label{eq:GLpsi1eqHarm1}
\end{eqnarray}
The solution of \refeq{GLpsi1eqHarm0} is proportional to the unperturbed eigenfunction $\overline{\dirmod}_1(x) = C \dirmod_0 $ while the solution to \refeq{GLpsi1eqHarm1} is unique as the equation is non-singular.
The general solution of \refeq{GLpsi1eq} is therefore of the form given in \refsec{Sens2} with 
$\dirmod_1^\perp = \widetilde{\dirmod}_1(x) \sin( \beta z)$.
Considering that $\linop_1 \dirmod_1 = -U_0 \sin^2( \beta z) \partial_x \widetilde{\dirmod}_1(x)$, 
\refeq{sens2} reduces to 
% \begin{equation} 
$\eigval_2 = { \bra \adjmod_0(x) , -U_0 \sin^2( \beta z) \partial_x \widetilde{\dirmod}_1(x) \ket /  \bra \adjmod_0(x) , \dirmod_0 (x)\ket }$ and therefore
 \begin{equation} 
 \eigval_2 = - \frac{U_0}{2} { \int_{x_a}^{x_b} {\adjmod_0}^*(x)\, \partial_x \widetilde{\dirmod}_1(x)\, dx \over  
                      \int_{x_a}^{x_b} {\adjmod_0}^*(x)\, \dirmod_0(x)\, dx}
 \label{eq:GLsens2}
\end{equation}
where the spanwise integration $\int_0^{2 \pi / \beta} \sin^2(\beta z) = 1/2$ has been used.
This result will be applied to a non-parallel and a parallel case in the next section.

%%%%%%%%%%%%%%%%%%%%%%%%%%%%%%%%%%%%%%%%%%%%%%%%%%%%%%%%%%%%%%%%%%%%%%%%%%%%%%%
\section{Stabilizing effect of spanwise modulations of the advection velocity}
\label{sec:Results}

\begin{figure}
\centering
%   \centerline{
\includegraphics[width=0.8\columnwidth]{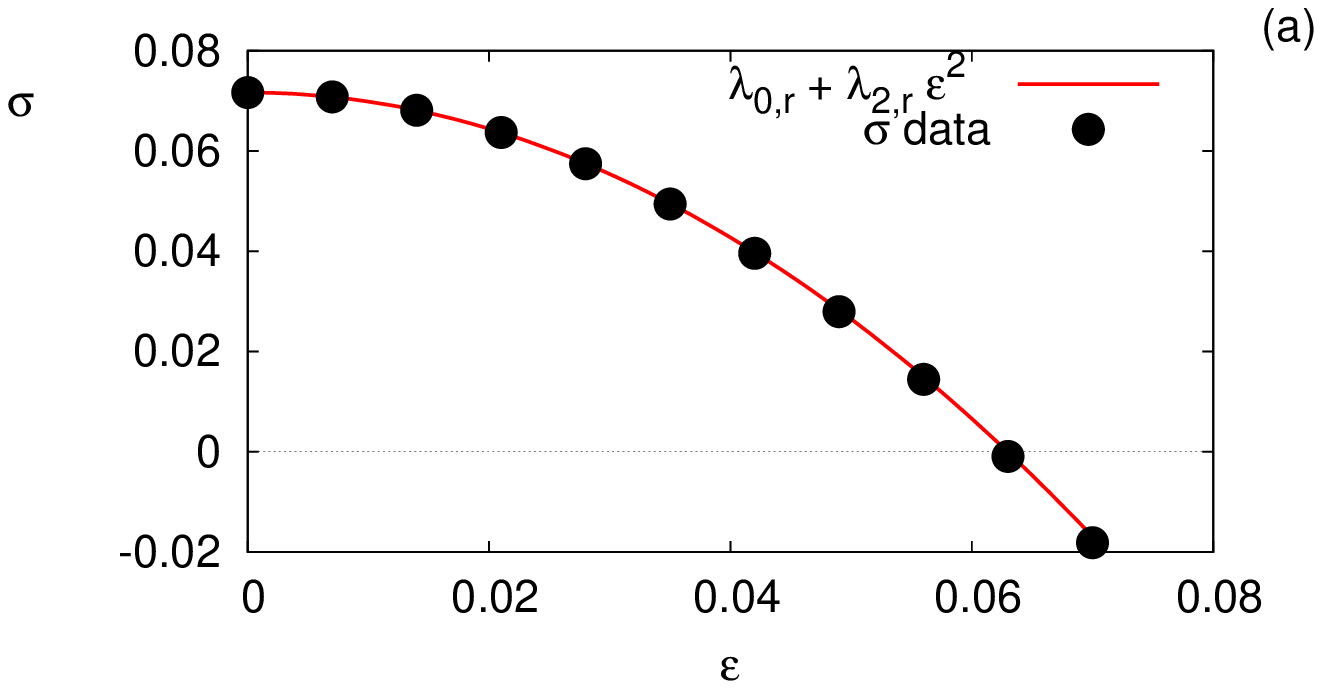}\\
\includegraphics[width=0.8\columnwidth]{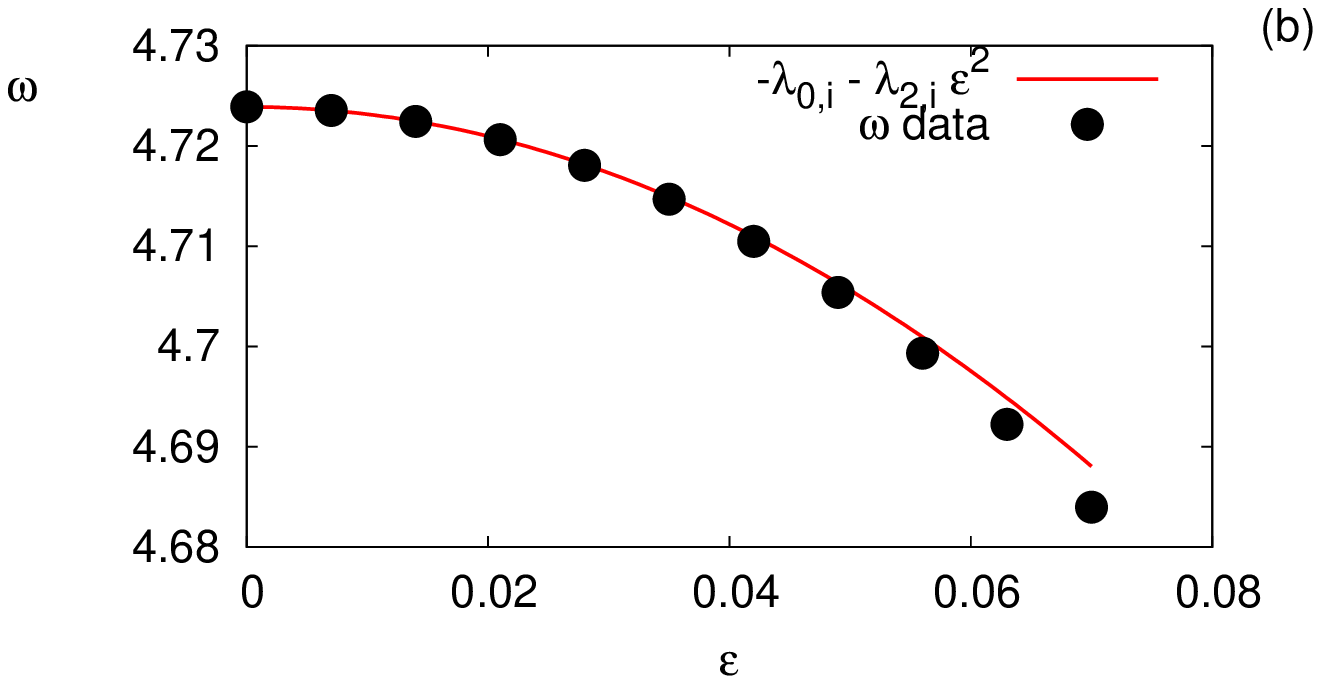}
%}
  \vspace*{-1mm}
  \caption{Dependence of the unstable global mode growth rate $\sigma$ (panel $a$) and frequency $\omega$ (panel $b$) on the amplitude $\pertamp$ of spanwise periodic modulations of the advection velocity
$U(z)  = U_0 \left( 1 + \pertamp \sin \beta z \right)$. 
Symbols (filled circles): results of a direct numerical computation of the eigenvalues.
Line (red online): prediction of the second order sensitivity analysis.
The global mode is stabilized for $\pertamp \gtrsim 6.3$\%.
} 
\label{fig:Sens2}
\end{figure}

\begin{figure}
%\centering
%   \centerline{
%\hspace{-4mm} 
\includegraphics[width=0.9\columnwidth]{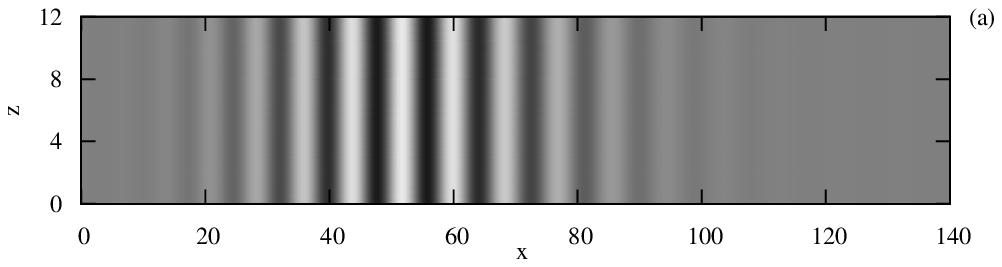}
%  \vspace*{-3mm}
%\hspace{-4mm} 
\includegraphics[width=0.9\columnwidth]{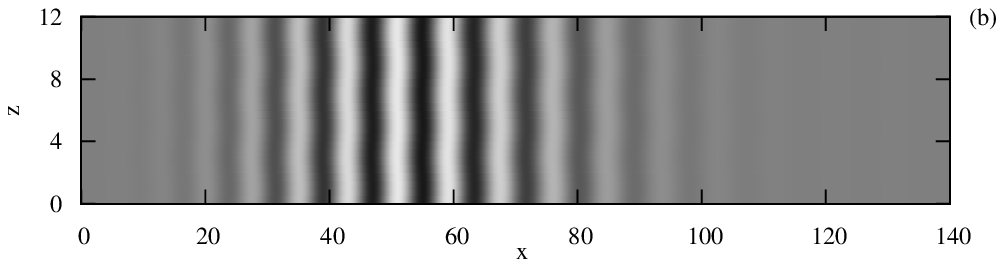}
 %\vspace*{-3mm}
%\hspace{-4mm} 
\includegraphics[width=0.9\columnwidth]{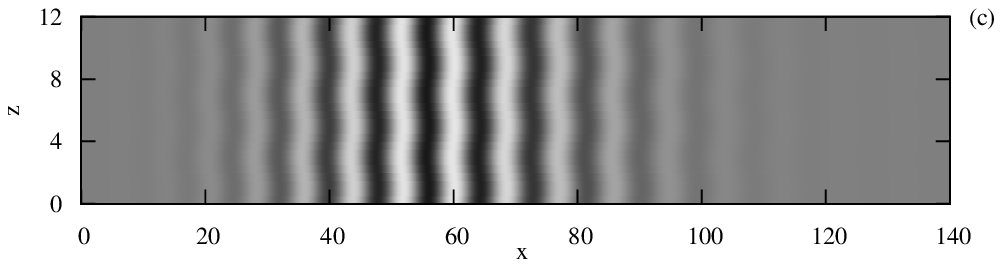}
%}
 \vspace*{-1mm}
\caption{Real part of the leading global mode $\dirmod$ of the linear Ginzburg-Landau operator with spanwise uniform advection velocity $U$ (panel $a$, reference 2D case with  $\pertamp=0$) and with increasing spanwise modulations of $U$ corresponding to a moderate growth rate reduction ($\pertamp=0.021$, panel $b$) and to an almost neutral global mode ($\pertamp=0.063$, panel $c$). The eigenfunctions are represented over three wavelengths in the spanwise direction in grey scale with white corresponding to the maximum positive value of $Re(\dirmod)$ and black to the minimum negative value. }
\label{fig:3Dmodes}
\end{figure}

\begin{figure}
\centering
%   \centerline{
\includegraphics[width=0.98\columnwidth]{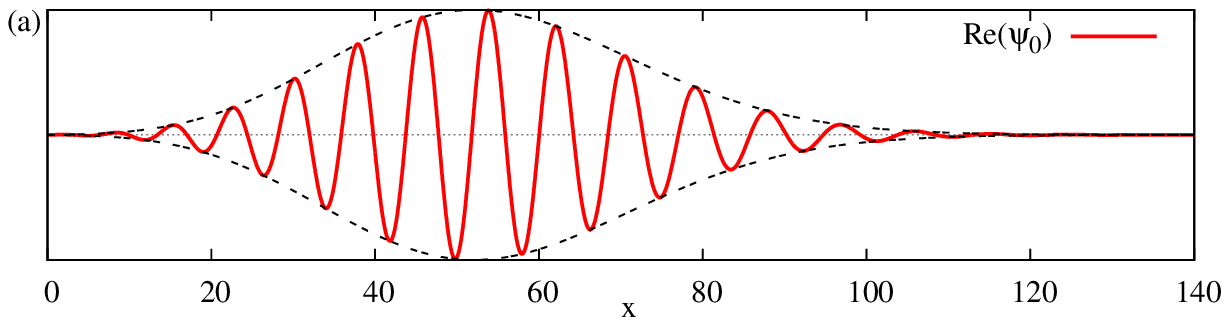}
\vspace{-2mm}

\includegraphics[width=0.98\columnwidth]{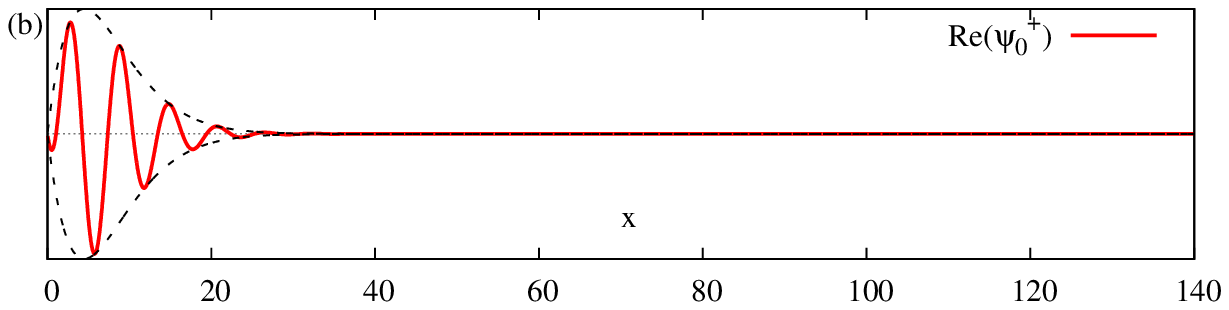}
\vspace{-2mm}

\includegraphics[width=0.98\columnwidth]{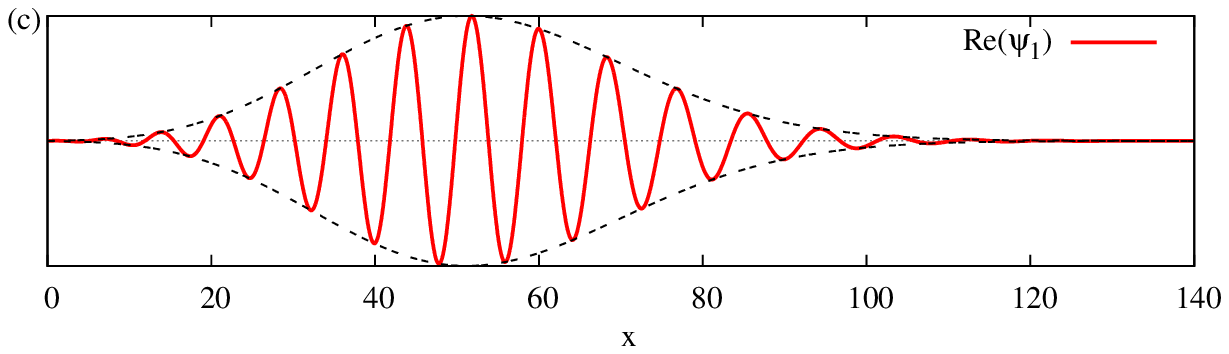}
\vspace{-1mm}
%}
  \vspace*{-1mm}
  \caption{Direct (panel $a$) and adjoint (panel $b$) global modes $\dirmod_0$ and $\adjmod_0$ of the 2D (spanwise uniform) reference Ginzburg-Landau model. The first-harmonic Fourier mode of the first-order correction to the global mode $\widetilde{\dirmod}_1$ is also shown (panel $c$). 
The carrier waves (real part of the functions) are reported with solid lines (red online) and the corresponding envelope (absolute value of the functions) with dashed lines (black) in normalized arbitrary units. 
The shape of $|\widetilde{\dirmod}_1|$ is almost identical to that of $|\dirmod_0|$.} 
\label{fig:Modes}
\end{figure}

%------------------------------------------------------------------------------
\subsection{Stabilization of the global instability}
\label{sec:Global}

%...............................................................................
%\subsubsection{The non-parallel generalized Ginzburg-Landau model and its global stability}
%\label{sec:Abso}

Let us first consider the non-parallel Ginzburg-Landau equation with the linear law $\mu(x) = \mu_{max} - \xi x $, with $\xi > 0$, $x \in [0,\infty[$ and $\glvar(x=0)=0$ which has been used to mimic bluff body wake dynamics near the critical Reynolds number \cite{Chomaz1988,Hunt1991}. 
With the chosen $\mu(x)$ for the spanwise-uniform (2D, no $z$ dependence) reference case,
eigenvalues and eigenfunctions can be computed analytically  \cite{Chomaz1988}:
\begin{eqnarray}
\eigval_0^{(n)} &=& \mu_{max}-\left(U_0^2 / 4 \gamma \right)
              +\left(\gamma \xi^2\right)^{1/3} \zeta_n, \\
\dirmod_0^{(n)}(x) &=& Ai \left(\zeta_n + \left[\xi / \gamma \right]^{1/3} x \right)
\label{eq:2Dspec}
\end{eqnarray}
where $Ai$ is the usual Airy function \cite{Abramowitz}, $\zeta_n$ is its $n-th$ zero  ($n=1,2, \ldots$) and $U_0$ the spanwise uniform value of $U$.
The global stability is determined by the sign of the real part of the leading eigenvalue $\eigval_0^{(1)}$
with the system globally unstable if $\mu_{max} > \mu_c$, where
%\begin{equation}
$\mu_c = U_0^2/ 4 |\gamma|^2 - \zeta_1 |\xi^2 \gamma|^{1/3} \cos[((1/3) \tan^{-1}(c_d))]$.
As $\zeta_1 < 0$, it is seen how the onset of the global instability always occurs in the presence of a finite region of local absolute instability that appears as soon as $\mu > U_0^2/ 4 |\gamma|^2$ (see \cite{Chomaz1988,Huerre1990}, but also next subsection).

%\subsubsection{Stabilizing effect of spanwise modulaitons of the adveciton velocity}
%\label{sec:Abso}

The reference non-parallel 2D case is chosen with the specific values $U_0=10$, $c_d=-5$ and $\xi=0.01$ which are similar to ones previously used to mimic globally unstable wakes  \cite{Chomaz1988}.
A globally unstable case with a single unstable eigenvalue $\lambda_0$ when $U=U_0$ is obtained by choosing 
$\mu_{max} = 1.2$.

We first verify if spanwise periodic $U$ modulations of spanwise wavelength $\lambda_z = 2 \pi / \beta$ have or not a stabilizing effect  by numerically computing the (global) eigenvalues and eigenfunctions of $\linop$, just as in recent studies of wake stabilization based on the full Navier-Stokes equations \cite{DelGuercio2014b,DelGuercio2014c}. Numerical details are given in Appendix~\ref{sec:Num}.
% This is done by first discretizing the linear equation on a streamwise-spanwise grid and by then computing the eigenvalues by iterative subspace iteration methods (see Appendix~\ref{sec:Num} for details).
The results obtained for $\lambda_z = 4$ are reported in \reffig{Sens2}, where the dependence of the growth rate $\sigma$ and of the frequency $\omega$ of the unstable global mode with eigenvalue $\lambda=\sigma  - i \omega$ are reported with filled circles for selected streaks amplitudes ranging from $\pertamp=0$ to $\pertamp=0.07$.
From \reffig{Sens2} it is seen that increasing amplitudes $\pertamp$ of the spanwise modulation of $U$ reduce both the growth rate of the global mode and its oscillation frequency. 
A critical amplitude exists $\pertamp_c \approx 6.3$\%, above which the global mode is completely stabilized.  
The variation of $\eigval$ begins with zero tangent near $\pertamp=0$ and assumes an almost-quadratic dependence for larger values of $\pertamp$. 
A similar trend is observed for the oscillation frequency $\omega$.
The eigenfunctions $\dirmod(y,z)$ become increasingly threedimensional as $\pertamp$ is increased, as shown in \reffig{3Dmodes}.
All these results reproduce well those found in non-parallel wakes \cite{DelGuercio2014b,DelGuercio2014c} confirming that the present model equation study captures essential features of the streaks stabilizing effect on the global instability of non-parallel wakes.

%\subsubsection{Accuracy of the second order structural sensitivity analysis}
%\label{sec:Abso}
As a second step of the analysis, the accuracy of the predictions of the second order sensitivity analysis are assessed by comparison with the direct eigenvalue computations. This is an important issue, since the applicability of this asymptotic analysis to moderately large amplitudes of the perturbations can not be a priori postulated.
To this end the formulation detailed in \refsec{Sens2CGL} is applied to the present non-parallel case.
The leading direct and adjoint eigenfunctions $\dirmod_0(x)$ and $\adjmod_0(x)$ of the unperturbed problem are  reported  in \reffig{Modes}.
The direct mode corresponds to a carrier wave modulated in the streamwise direction that starting from its zero value enforced in $x=0$, increases its amplitude in the region of local instability to then relax to zero downstream in  the locally stable region of the flow \cite{Chomaz1988,Hunt1991}.
The adjoint mode is localized more upstream than the direct mode, as expected from the inversion of the advection speed in the adjoint operator.
In order to obtain $\widetilde{\dirmod}_1(x)$, Equations~(\ref{eq:GLpsi1eqHarm0})  and (\ref{eq:GLpsi1eqHarm1}) are discretized and then solved  as detailed in Appendix~\ref{sec:Num}.
The solution is reported in \reffig{Modes}$c$, from which it is seen that for this specific case, the  envelope of $\widetilde{\dirmod}_1$ is almost coincident with that of $\dirmod_0$.
The second order sensitivity is then retrieved from \refeq{GLsens2}, with $x_a=0$ and $x_b=L_x$, with $L_x$ large enough, as detailed in Appendix~\ref{sec:Num}.
For the considered parameter values, it is found $\eigval_2=-18.08 + 7.32\, i$.
Using this value, the prediction of the second order sensitivity analysis $\eigval = \eigval_0 + \eigval_2 \pertamp^2$, reported in \reffig{Sens2}, are found to closely match the results issued from the much more expensive computation of the eigenvalues, also reported in the same figure.
A small loss of accuracy is observed only for large values $\pertamp$ and only for the fit to $\omega$.

The robustness of these results has been tested by verifying that the result do not qualitatively change when different values of the parameters are used and when the shape of $\mu(x)$ is changed to quadratic as in e.g.~\cite{Cossu1997c}.

The stabilizing effect of streaks can therefore be well reproduced in the complex Ginzburg-Landau equation by simply modulating the advection velocity in the spanwise direction. This effect is therefore not limited to a particular vortex dynamics mechanism active in wakes and is not limited to the  Navier-Stokes equation. The found effect is much more general nature and could probably be exploited in different physical contexts which can be described by the complex Ginzburg-Landau equation.

%------------------------------------------------------------------------------
\subsection{Stabilization of the local absolute instability}
\label{sec:Abso}

%  Having established that with the sole spanwise variation of wave the advection velocity $U$, the main features of the streaks stabilizing action on wakes are retrieved, we now want to better understand to which mechanism this stabilization can be attributed.
In studies of wakes, it has been established that the stabilization of the global unstable mode is associated to the weakening of the absolute instability in the near wake. 
The absolute or convective nature of an instability is a local concept that must is defined for parallel flows. 
We therefore consider the effect of spanwise variations of the advection velocity on the `parallel' Ginzburg-Landau model where the coefficient $\mu$ is constant, i.e. it does not change with the streamwise coordinate $x$.

%...............................................................................
%\subsubsection{Local absolute instability of the uncontrolled parallel flow}
%\label{sec:Abso}

Let us first  briefly summarize the well known main results obtained in the parallel 2D case (see e.g. \cite{Chomaz1988,Huerre1990,Chomaz2005}).
In this case, the eigenfunctions are $\dirmod_0=e^{i \alpha x}$ with associated eigenvalue
$\eigval_0 = \mu - i\,\alpha\,U_0  - \gamma\,\alpha^2$, where $\alpha$ is the streamwise wavenumber.
% The adjoint eigenfunction coincides with the direct.
The absolute growth rate is the one that would be observed at a fixed location for large times.
It can be obtained by first computing the complex wavenumber $\alpha_{0,s}$ corresponding to zero group velocity
%$d \eigval_0 / d \alpha = - i\,U_0  - 2 \gamma\,\alpha = 0$ 
$d \eigval_0 / d \alpha = 0 $ which gives  
$\alpha_{0,s} = - i\,U_0  / 2 \gamma$ in correspondence to which  
$\eigval_{0,s} = \mu - U_0^2 / 4 \gamma$. 
The absolute growth rate $\sigma_0$ that would be measured at a fixed streamwise station for large times is the real part of $\eigval_{0,s}$ i.e. 
$\sigma_0 = \mu_r - \gamma_r U_0^2 / 4 |\gamma|^2$, where the subscript $r$ denotes the real part.
In the parallel case therefore, as it is well known, linear instability sets in as soon as $\mu_r > 0$, but the instability becomes absolute and grows also in place, only for $\mu_r > \gamma_r U_0^2 / 4 |\gamma|^2$.

%...............................................................................
%\subsubsection{Structural sensitivity of the local absolute growth rate}
%\label{sec:Abso}

Let us now examine the influence of the spanwise modulation of $U$ on the absolute instability.
%  in the parallel case.
Making use of the streamwise Fourier transform to solve 
\refeq{GLpsi1eqHarm1} with $\dirmod_0=e^{i \alpha x}$, it is readily found that $\widetilde{\dirmod}_1=- (i \alpha U_0 / \beta^2)\,e^{i \alpha x}$.
From \refeq{GLsens2}, with integrals evaluated between $x_a=0$ and $x_b=2 \pi / \alpha$ it comes that  
$\eigval_2= - U_0^2 \alpha^2 /2 \beta^2$, which represents the action of an additional diffusive term.
In the presence of streaks of amplitude $\pertamp$, therefore
% \begin{eqnarray}
$\eigval= \mu - i\,\alpha\,U_0  - \gammaeff\,\alpha^2$ with 
$\gammaeff = \gamma + \pertamp^2 U_0^2 / 2 \beta^2$.
The value of the (streamwise) absolute growth rate is retrieved as in the reference case but with $\gammaeff$ replacing $\gamma$:
$\eigval_{s} = - i\,U_0  / 2 \gammaeff$.
Expanding $1 / \gammaeff$ in terms of $\pertamp$ up to second order it is readily found that
$\eigval_{s} \sim \eigval_{0,s} + \eigval_{2,s} \pertamp^2$  with $\eigval_{2,s} = {U_0^4}/{ 8 \beta^2 \gamma^2}$, and, taking the real part 
$\sigma = \sigma_0 + \sigma_2 \,\pertamp^2$. The second order sensitivity of the absolute growth rate to spanwise periodic modulations of the advection velocity is therefore 
$\sigma_2= - {(c_d^2-1) U_0^4}/{ 8 \beta^2 (1 + c_d^2)^2}$.
The effect is stabilizing as long as $c_d > 1$.

%...............................................................................
%\subsubsection{Local interpretation of the stabilization of the global instability}
%\label{sec:Abso}

The stabilization of the global instability described in \refsec{Global} can now be interpreted in terms of local absolute growth rate stabilization.
In the considered case $\mu(x)=\mu_{max} - 0.01\,x$ and, for the chosen parameters, the critical value of $\mu_{max}$ for the global instability is $\mu_c \approx 1.13$.
At the onset of the global instability,  $\mu_{max}=\mu_c$ and  the maximum (uncontrolled) absolute growth rate is $\sigma_0(x=0)=\mu_c - U_0^2 / 4 |\gamma|^2 \approx 0.17$ with and the absolutely unstable region extending to $x \approx 17$.
The globally unstable reference case of \refsec{Global} was chosen with $\mu_{max}=1.2$, for which the maximum absolute growth rate is $\approx 0.24$ and the absolute instability region extends up to $x \approx 24$.
In the presence of spanwise modulations of the advection velocity of amplitude $\pertamp$, the absolute growth rate is given by 
$\sigma(x) = \mu(x) - 0.96 - 17.99\, \pertamp^2$.
We have seen that the global instability is stabilized for control amplitudes $\pertamp \gtrsim 0.063$.
At the critical value $\pertamp \approx 0.063$, the maximum absolute growth has been reduced to $\approx 0.17$ with the absolute instability region extending only to $x \approx 17$. 
At the restabilization, therefore, the local absolute growth rates have been reduced to the value they would have had at $\mu_{max}=\mu_c$.
The suppression of the global instability by the spanwise $U$ modulations must therefore be ascribed to the reduction of the local absolute instability growth rate.

%%%%%%%%%%%%%%%%%%%%%%%%%%%%%%%%%%%%%%%%%%%%%%%%%%%%%%%%%%%%%%%%%%%%%%%%%%%%%%%
\section{Discussion and conclusions}
\label{sec:Concl}

The goal of the present study was to understand the origin of streaks stabilizing effect on the global and local absolute instabilities in bluff-body wakes revealed by recent studies \cite{Hwang2013,DelGuercio2014,DelGuercio2014b,DelGuercio2014c}.
To this end it has been investigated if this stabilizing effect could be reproduced in the generalized complex Ginzburg-Landau equation that has been previously used to elucidate the dynamics of local and global instabilities in open flows \cite{Chomaz1988,Huerre1990,Hunt1991,Cossu1997c,Chomaz2005}.
As the streaks are associated to spanwise modulations of the wake streamwise velocity profile, it has been guessed that their leading effect is to induce a spanwise modulation of the advection velocity of unstable waves. It has then been verified if such a modulation has or not a stabilizing effect on global and local instabilities.

As a first result, it has been shown that spanwise modulations of the advection velocity have a stabilizing effect on the global instability of the non-parallel complex Ginzburg-Landau equation with coefficients chosen so as to mimic the global instability of non-parallel wakes. 
The growth rate of the global mode has been shown to decrease quadratically with the amplitude of the modulations and the global instability has been suppressed for large enough modulation amplitudes.
These results are in complete agreement with what found in non-parallel wakes \cite{DelGuercio2014b,DelGuercio2014c} and show that the simple spanwise modulation of the advection velocity plays a key role in the stabilization mechanism. 
As no vorticity is defined in the Ginzburg-Landau equation, this key stabilizing mechanism is simpler and of more general nature than explanations based on the vortex dynamics of wake vortices in the presence of 3D modulations.  
This stabilizing effect is also more general in the sense that it can probably be applied to other physical systems described by the complex Ginzburg-Landau equation.

A second order structural sensitivity analysis has then been shown to provide accurate predictions of the variation of the growth rate of the unstable global mode with respect to the amplitude of the advection velocity modulations.  
This second order sensitivity analysis has then been applied to compute the influence of these modulations on the eigenvalues and on the absolute growth rate in the parallel case. 
It has been shown than the leading order effect of the advection velocity modulations is to alter the wave  diffusivity in the local dispersion relation. 
This altered diffusivity is shown to be associated to a decrease of the absolute growth rate.
We believe that this mechanism lies at the core of the absolute growth rate reductions observed in parallel wakes \cite{Hwang2013,DelGuercio2014}. 
A simple local stability analysis of the global mode stabilization obtained in the non-parallel case shows that at criticality the spanwise advection velocity modulations have reduced the pocket of local absolute instability to the same level that it would have had by reducing the global bifurcation parameter $\mu_{max}$ to its critical value $\mu_c$. 
This confirms that the suppression of the global instability is induced by the reduction of the local absolute instability.

A side-result of this investigation is the validation of the effectiveness and accuracy of the second order structural sensitivity analysis in predicting eigenvalues variations induced by spanwise periodic basic flow modulations. 
Additional effects can be easily investigated with this methodology. 
We have, for example, found that spanwise periodic modulations of $\mu$ have a slightly destabilizing effect which is less important than the stabilizing effect of $U_1$. The $U$ modulation is therefore the only relevant mechanism in the stabilization.
Also, it has been easy to verify that conclusions similar to the ones we have described, are reached when instead of the linear variation of $\mu(x)$ with $x$, a quadratic variation is considered (with the same parameters considered in Ref.~\onlinecite{Cossu1997c}) and/or when a strong non-parallelism of the basic flow is mimicked increasing $\xi$ to $O(1)$. 
The second order sensitivity analysis, which does not seem to have never been used with the Navier-Stokes equations, is also computationally much less expensive than the repeated eigenvalues computations for the full 3D problem. The application of this analysis to Navier-Stokes computations of the stabilization of bluff body wakes is currently the object of intensive effort.

% DISCUSS:
% 
% UNPHYSICAL DEPENDENCE ON 1/BETA2... NOT TRUE THAT VERY LONG WAVES ARE MORE STABILIZING... PROBABLY SPANWISE DIFFUSION NEEEDS TO BE IMPROVED...;

~ 

%%%%%%%%%%%%%%%%%%%%%%%%%%%%%%%%%%%%%%%%%%%%%%%%%%%%%%%%%%%%%%%%%%%%%%%%%%%%%%%
\appendix

\section{Numerical methods}
\label{sec:Num}

The non-parallel generalized complex Ginzburg-Landau equation has been discretized, first to directly compute the global eigenvalues and modes and then to perform the second order structural sensitivity analysis. 

For the direct eigenvalues computation, the equation is discretized using second order accurate finite differences in $x$ and Fourier expansions in $z$ on a grid of $N_x \times N_z$ regularly spaced points in the domain $[0,L_x] \times [0 , L_z]$. 
It has been verified that the local convective or absolute nature of the instability is not modified by numerical effects \cite{Cossu1998}.
Some preliminary tests have shown that with $L_x=140$ and $N_x=4000$ the leading eigenvalue of the 2D case is retrieved with a precision superior to 1\%. 
A similar precision is attained for the considered range of $\pertamp$, by setting $N_z=8$ for $L_z= 4$. The eigenvalues of the discretized version of $\linop$ are then computed for each selected value of $\pertamp$ using the {\sl ArPack} suite \cite{ARPACK}.

For the structural sensitivity computations, the sensitivity equations are discretized using the same numerical schemes and parameters used to discretize $\linop$,  except for the fact that no grid is necessary in the spanwise direction (which greatly reduces the computational cost). 
The unperturbed (2D) problem is preliminary solved and gives 
$\eigval_0$, $\dirmod_0$ and $\adjmod_0$ which are positively  cross-checked with the available exact solution. 
$\widetilde{\dirmod}_1(x)$ is computed by solving the discretized version of \refeq{GLpsi1eqHarm1}. 
With the used second order accurate finite difference formul\ae, the solution of \refeq{GLpsi1eqHarm1} is quickly obtained making use of standard fast algorithms for the inversion of tri-diagonal banded matrices.

%%%%%%%%%%%%%%%%%%%%%%%%%%%%%%%%%%%%%%%%%%%%%%%%%%%%%%%%%%%%%%%%%%%%%%%%%%%%%%%%%%%%%%%%%%%%%%
%CHANGE BIBSTYLE BELOW

%\bibliographystyle{aipnum4-1.bst}
%\bibliography{carlo3}

%merlin.mbs apsrev4-1.bst 2010-07-25 4.21a (PWD, AO, DPC) hacked
%Control: key (0)
%Control: author (8) initials jnrlst
%Control: editor formatted (1) identically to author
%Control: production of article title (-1) disabled
%Control: page (0) single
%Control: year (1) truncated
%Control: production of eprint (0) enabled
\newcommand{\noopsort}[1]{} \newcommand{\printfirst}[2]{#1}
  \newcommand{\singleletter}[1]{#1} \newcommand{\switchargs}[2]{#2#1}
%

% Produces the bibliography via BibTeX.
%%%%%%%%%%%%%%%%%%%%%%%%%%%%%%%%%%%%%%%%%%%%%%%%%%%%%%%%%%%%%%%%%%%%%%%%%%%%%%%%%%%%%%%%%%%%%%
%%%%%%%%%%%%%%%%%%%%%%%%%%%%%%%%%%%%%%%%%%%%%%%%%%%%%%%%%%%%%%%%%%%%%%%%%%%%%%%%%%%%%%%%%%%%%%
\end{document}